\documentclass[useAMS,usenatbib]{mn2e}
\usepackage[pdftex]{graphicx}
\usepackage{subfigure}
\graphicspath{{MNRAS_fig/}}	
\pdfoutput=1

\title[Optical turbulence - SAAO Sutherland site]{Optical turbulence characterization at the SAAO Sutherland site}

\author[L. Catala et al.]
{\parbox{\textwidth}{L. Catala$^{1,2}$\thanks{E-mail:lcc@saao.ac.za},
S. M. Crawford$^{1}$,
D.A.H. Buckley$^{3,1}$,
T. Pickering$^{3,1}$,
R. Wilson$^{4}$,
T. Butterley$^{4}$,
H. Shepherd$^{4}$,
F. Marang$^{1,3}$,
P. Matshaya$^{1,3}$ and
C. Fourie$^{1}$}\vspace{0.4cm}\\
\parbox{\textwidth}{$^{1}$South African Astronomical Observatory, Observatory Road, Observatory 7935, South Africa\\
$^{2}$University of Cape Town, Private Bag X3, Rondebosch 7701, South Africa\\
$^{3}$Southern African Large Telescope, P.O. Box 9,Observatory 7935, South Africa\\
$^{4}$Department of Physics, Center for Advanced Instrumentation, University of Durham, South Road, Durham DH1 3LE, England}}

\begin{document}

\date{Accepted 2013 August 21. Received 2013 August 19; in original form 2013 February 19}

\pagerange{\pageref{firstpage}--\pageref{lastpage}} \pubyear{2013}

\maketitle

\label{firstpage}
\begin{abstract}
	
  We present results from the first year of a campaign to
  characterize and monitor the optical turbulence profile at the SAAO
  Sutherland observing station in South Africa. A MASS-DIMM (Multi
  Aperture Scintillation Sensor - Differential Image Motion Monitor)
  was commissioned in March 2010 to provide continuous monitoring
  of the seeing conditions. Over the first month of the campaign, a
  SLODAR (SLOpe Detection And Ranging) from Durham University
  was also installed allowing an independent verification of the
  performance of the MASS-DIMM device. After the first year of data
  collection, the overall median seeing value is found to be 1.32"
  as measured at ground level. The ground layer which includes all 
  layers below 1 km accounts for 84\% of the turbulence, while the 
  free atmosphere above 1 km accounts for 16\% with a median value 
  of 0.41". The median isoplanatic angle value is 1.92", which is 
  similar to other major astronomical sites. The median coherence 
  time, calculated from corrected MASS measurements, is 2.85 ms. 
  The seeing conditions at the site do show a strong correlation 
  with wind direction with bad seeing conditions being associated 
  with winds from the South-East.

\end{abstract}
\section{Introduction}

The degradation of astronomical image quality due to the distortion of
light from atmospheric turbulence has been a long-known problem to
astronomers \citep{b2,b3}. Temperature and density gradients
induce variations in the refractive index. As explained in \citet{b1}, the atmospheric
turbulence arises from different layers in the atmosphere and is well
describe by a Kolmogorov model. Turbulence is
particularly evident in regions associated with wind shear between
layers \citep{b2}. The strength of the turbulence in the different
layers is commonly characterized by the index of refraction structure constant ($C_N^2(h)$),
which is dependent on the altitude, $h$.  Parameters used to
characterize the optical turbulence depend on the cumulative effect
from the turbulence from all layers in the atmosphere and can be
calculated by integrating the $C_N^2$ profile along the path.  These parameters
include the coherence length or Fried Parameter ($r_0$), the coherence
time ($\tau_0$), and the isoplanatic angle ($\theta_0$). In terms of
image quality, the astronomical seeing ($\epsilon_0$), assuming a Kolmogorov turbulence model 
(ie. an infinite outer scale), is generally defined as
the full width at half maximum (FWHM) of a long exposure image point
spread function (PSF) and is inversely proportional to the Fried
parameter such that FWHM = 0.98$\lambda /r_{0}$\citep{b1}.

Over the last 20 years, atmospheric site characterization and seeing
monitoring for all major astronomical observatories has come into
common usage \citep{b4,b5,b6}. Indeed, seeing quality is one of the
main critical parameter for site selection \citep{b7,b8}. In addition,
continuous monitoring of the seeing is critical for optimizing
observing schedules, especially for queue-scheduled observations
\citep{b9}. With the advent of adaptive optics (AO) systems
\citep{b10,b11}, one can now compensate for the image distortion due
to atmospheric turbulence. Such systems rely on determining the
turbulence properties of a site, and extensive characterization of the
atmospheric turbulence are necessary.

The Sutherland observing station of the South African Astronomical Observatory
(SAAO) was first quantitatively characterized in 1992 from photographic seeing measurements, 
and then later, these were supplemented by DIMM measurements from 1994 to 2000
\citep{b12,b13}. Since then, very little has been done in terms of optical site 
quality studies. In 2010, we initiated a comprehensive site characterization 
campaign to improve the scheduling of the Southern African Large Telescope (SALT) 
observations and provide the basis for a feasibility study of an AO system for the SALT.

In this paper we present the first year of seeing data. In \S 2, the
main features of the site along with results from previous studies are
presented. The Sutherland site monitoring instrument setup, data
processing and analysis, and observing periods for the present study
are described in \S 3. We describe the data analysis method and
discuss data consistency in \S 4. The results are presented in \S 5
and discussed in \S 6. Finally, we conclude by summarizing the main
results and future objectives for the Sutherland site characterization.

\section{The SAAO Sutherland site}

\begin{figure*}
\includegraphics[scale=0.4]{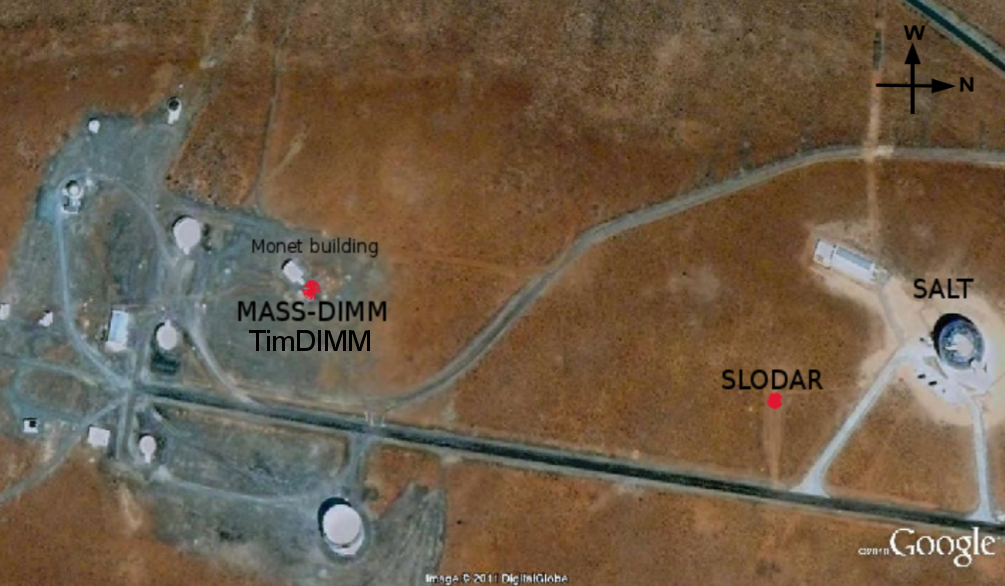}
  \caption{Sutherland site and the seeing monitoring instruments location}
  \label{site}
\end{figure*}
\subsection{Location}

The Sutherland site, located 370 km North-East of Cape Town in the 
semi-arid Karoo region, was founded in 1972 by merging the Republic 
Observatory (Johannesburg) and the Radcliffe Observatory (Pretoria) 
with the Royal Observatory (Cape of Good Hope). It is at 32$^o$23' S 
latitude and 29$^o$49' E longitude. The Sutherland plateau, host of 
12 optical and infrared telescopes, has an altitude of 1768 m above sea level.

\subsection{Historical site characterization results}
\label{sec:site}

The first historical report of seeing conditions at Sutherland was
made by \citet{b40} and indicated 69\% of the nights had seeing better
than 0.8'' from visual observations\footnote{The seeing reported in
  \cite{b40} was converted to FWHM by \cite{b41}}. Seeing measurements
by \citet{b41}, also based on eye measurements made during slit-scanning
photometry of double stars observations between 1972 and 1994, reported
53\% of the nights had seeing better than 0.8".  Due to the
methodology used, it is difficult to evaluate the reliability of these
measurements to compare to later quantitative results given by dedicated seeing
monitors.
  
Seeing measurements were carried out using a trailed photographic technique 
from February 1992 to May 1993 by \citet{b13}. These results gave a median 
seeing of 1.23", but were subject to wind effects.
  
Differential Image Motion Monitor (DIMM, \citealt{b14,b15})
observations were first obtained from April 1994 to February 1998. 
The system was based on that of \citet{b44}. These results were reported in 
\citet{b12} and indicated that the median seeing value for
this period was 0.92" and the first and third quartiles were,
respectively, 0.74" and 1.16" \citep{b12}.  A site testing campaign, 
using two DIMM instruments, was carried out to select the final site 
for SALT, from December 1998 to March 2000, and showed similar
statistics with a median value of 0.95", and first and third quartiles
of 0.79" and 1.18" \citep{b12}.

The seeing conditions measured over the period from April 1994 to
February 1998 showed significant correlation with the wind speed and
direction \citep{b12}. South-Easterly winds brought poorer conditions while
the best seeing value were observed for westerly winds. Also, worse
seeing values tended to be associated with higher wind speed.  The
seeing conditions showed a very small dependence on the seasons.
Observations during the SALT site testing campaign verified these
correlations \citep{b12}.

Considering the topography together with wind speed and directions at
different altitudes, the main contributions to the turbulence were
expected to be the surface layer up to 1 km, a wind shear layer around
3 km, and one between 10 and 12 km due to the jet stream \citep{b12}. 
In addition, a 40 min scan from a classical SCIDAR (SCIntillation
Detection And Ranging, Vernin \& Roddier 1973) and
microthermal measurements from a 30 m mast gave
more information about the turbulence profile. The SCIDAR
scan showed 3 strong layers at the ground level, around 3 km and 12 km.
Based on the microthermal data,
\citet{b12} found that the first 30 m of the ground layer (GL)
were contributing only 7.3\% of the overall turbulence with
a seeing of 0.19".


\section{The Sutherland site monitoring setup}

Over the past few decades several different methods have been developed to accurately measure atmospheric turbulence and turbulence profiles (TP).  One of the more well-established methods combines a Multi Aperture Scintillation Sensor (MASS) and a Differential Image Motion Monitor (DIMM) into a single MASS-DIMM device \citep{b23}. Such a device was installed and commissioned at the SAAO site in Sutherland in March 2010 with the objective of providing fully-automated, continuous seeing and turbulence profile measurements.  During the year 2010 a SLOpe Detection And Ranging (SLODAR) instrument \footnote{on loan from Durham University} \citep{b20} and a secondary DIMM instrument (hereafter referred to as TimDIMM) were also used alongside the MASS-DIMM.  Eventually, in February 2011, the secondary DIMM operation stopped and its camera and operating procedure were transferred to the DIMM channel of the MASS-DIMM.  The Sutherland instruments' working principles, setup, operation and data processing are described in this section along with a summary of the observing periods for the different instruments reported in this paper.

The location of the different seeing instruments on the plateau can be
seen on Fig.~\ref{site}. The SLODAR was located in its own enclosure, while the MASS-DIMM and TimDIMM shared the same enclosure hereafter referred to as the DIMM building. Nightly seeing measurements from the MASS-DIMM are reported at http://www.salt.ac.za/$\sim$seeing.  

\subsection{SLODAR}

\subsubsection{Working principle}

The SLODAR instrument is used to probe the ground layer (GL) profile
of the atmospheric turbulence. The method uses a double star target to
measure the spatial covariance of the slope of the wavefront phase
aberration, seen at ground level, of two different paths through the
atmosphere. In order to create two sets of spots, one for each star,
the system uses a Shack-Hartmann wavefront sensor,
mounted on a small aperture telescope. Using the spatial
cross-covariance of the measured centroids, one can infer the
turbulence profile \citep{b21}. The vertical resolution of the
profiles depends on both the angular separation of the double star and
its zenith distance. Aside from the GL profile, the SLODAR is also
using the DIMM technique, described later, to determine the overall
seeing value.

\subsubsection{Location and general setup}

The SLODAR was located approximately 100 m South-East from the SALT
building. It uses a Schmidt-Cassegrain Meade telescope with a 40 cm aperture 
diameter mounted on an equatorial pier at ground level. The system was 
in a dome enclosure protected by a 2 m high wind screen.

\subsubsection{Operation and control}

The SLODAR was set up by two of us and operated from the University of Durham. 
The Sutherland setup uses a 8x8 Shack-Hartmann wavefront sensor with 5 cm size sub-apertures. 
With the system being fully robotic, it was operated from Durham, for the first 4 hours of the night, on each night when weather conditions allowed. More details on the operation 
and control are given in \citet{b43}, which describes a similar setup of the instrument 
at Cerro Paranal. The higher vertical resolution of the Sutherland setup was achieved 
by using a CCD with a larger area that allowed a wider separation of the double stars. The 
resolution ranges from 55 to 80 meters, varying with star separation and elevation angle. 
Due to the wider separation, the maximum sensing altitude is around 500 m instead of 
1km on a regular setup.

\subsubsection{Data processing}
The SLODAR data were processed as extensively detailed in \citet{b43}. However the profile 
reconstruction model was modified and for the Sutherland data set is done using the "Mk II" 
analysis mentioned in \citet{b46}. The "Mk II" analysis fits an extra turbulence component at 
the ground with a non-Kolmogorov power spectrum in order to account for dome/tube seeing.
This component is also subtracted from the centroid autocovariance prior to fitting for r$_{0}$ 
to give an estimate of the total seeing that has been corrected for dome seeing.

\subsection{TimDIMM}

From July 2010 to January 2011 the TimDIMM was operated as a secondary DIMM instrument side by side with the MASS-DIMM. In February 2011, the CCD was migrated to the DIMM part of the MASS-DIMM to replace the SBIG-ST5, and the software was upgraded to handle the overall operation of the MASS-DIMM.

\subsubsection{DIMM working principle}

The DIMM is the most commonly used instrument for measuring the integrated seeing.  The method is based on the differential motion of two images of a single star. The system uses a two aperture mask at the entrance pupil of the telescope with one aperture fitted with a thin-wedge prism to create two images of the same star, but with rays that propagate through different paths in the atmosphere. This way the stellar images share a common mount and optical system which greatly suppresses systematic errors due to telescope vibration and wind buffering. The DIMM principle is presented and discussed in \citet{b15}. The variance of the differential image motion is calculated from a series of images and the seeing is directly deduced via its relationship with the variance \citep{b38}.

\subsubsection{Location and general setup}

The TimDIMM was located in the DIMM building next to the Monet-South telescope building, some 300 m South-East from the SALT. The system used a standard DIMM configuration \citep{b15} on a 25 cm LX200-GPS Meade telescope mounted on the standard Meade tripod with an Alt-Az fork mount. The entrance aperture of the telescope lied at approximately 1.5 m from the ground. The detector used was a IEEE1394 camera from Point Grey Research (specifically, a Grasshopper GRAS-03K2M) which is capable of frame rates up to 200 Hz (full frame 640x480 pixels with no binning). A frame rate of 330 frames/sec with 2x2 binning is used in normal TimDIMM operations for additional speed and sensitivity.

\subsubsection{Operation and control}

The system's pointing and tracking, as well as the real-time data analysis process are 
fully automated. The observing protocol is set so that the system selects targets with 
zenith angle less than 45$^o$ and a magnitude less than 3. The system acquires 10000 
frames to produce a seeing measurement. This provides a new seeing value every ~30 seconds 
when running at the standard rate of 330 frames/sec and a 1 to 3 ms exposure time.

\subsubsection{Data processing}

The TimDIMM data are processed as described in detail in \citet{b19}. The centroids of the stellar images are calculated via intensity-weighted moments within a 20x20 pixel box with a threshold set to 3$\sigma$ above the measured background noise in the image. Only pixels with signal above the threshold are included in the centroiding calculations.  The weighted variance of the longitudinal differential motion is calculated using the signal-to-noise ratios as weights. We are not currently calculating the transverse variance, which is more sensitive to bias from wind smoothing \citep{b45}. The formal uncertainties in the centroids are used to determine the variance due to measurement uncertainty.  This is then used to determine a corrected variance from which a seeing value is derived.   We are also not applying an exposure time correction.  With 1--3 ms exposures (depending on the star's brightness) and a 3 ms cadence we have found it largely unnecessary.  

\subsection{MASS-DIMM}
\subsubsection{MASS working principle}

The MASS instrument has been developed to provide a measurement of the
free atmosphere (FA) seeing as well as a low resolution profile of the
turbulence from 500m and above. Based on the well established
relationship between scintillation from a single star image and
atmospheric seeing \citep{b1}, it uses measurements of the
scintillation in four concentric pupil apertures via photomultiplier
tubes (PMTs) to recover the free atmosphere turbulence profile
\citep{b28}. The resolution and height of the measured C$_N^{2}$
values are set by the diameter of the different apertures acting as
spatial filters. Using both scintillation indices (SI) and
differential scintillation indices (DSI), the MASS, with its current
configuration, gives the C$_N^{2}$ values at 0.5, 1, 2, 4, 8 and 16 km
above the telescope. Moreover, the MASS also provides values for the
coherence time and the isoplanatic angle. A full description of the
device is given by \citep{b23}.

\subsubsection{MASS-DIMM configuration}
The MASS-DIMM combines two instruments in a single device, as described in \citet{b23}. This requires some modifications to the instrument's optical setup.  Instead of an aperture mask at the telescope's entrance pupil, an optical plate is positioned in the image pupil plane.  This plate consists of two mirrors that form the two apertures of the DIMM channel and a pupil segmentor unit made of four concentric mirrors that feeds the MASS channel.  An advantage of the MASS-DIMM is that it allows one to deduce the ground layer (GL) contribution to the turbulence by subtracting the free atmosphere (FA) seeing, given by the MASS channel, from the overall seeing measured by the DIMM channel \citep{b24}.

\subsubsection{Location and general setup}
The MASS-DIMM is located in the DIMM building next to the Monet-South
telescope building, some 300 m South-East from the SALT. The system
uses a 25 cm LX200GPS Meade telescope. From March 2010 to February
2011 it was mounted on the standard Meade tripod with an Alt-Az mount. Later, 
the mount was replaced by an Astro-Physics 900GTO equatorial
mount and bolted directly to a steel pier. In both configurations the entrance 
aperture of the telescope lies between 1 and 1,5 m from the ground.

\subsubsection{Operation and control}
The MASS-DIMM in its first configuration (February 2010 to January 2011) was operated manually.  The tracking was handled by the standard Meade Autostar II control system. The MASS data were handled by the Turbina 2.06 software \citep{b18}. The DIMM imaging was done by an SBIG-ST5 CCD and the data were processed by the RoboDIMM software \citep{b19}. In January 2011, the alt-az mount was replaced by the equatorial one. From February 2011 onwards, the Point Grey IEEE1394 camera was migrated over from TimDIMM and replaced the ST5 on the MASS-DIMM. The TimDIMM software was updated to manage and control the entire MASS-DIMM system, including DIMM data acquisition and processing, handling telescope pointing and tracking, conducting the initialization and remote operation of the Turbina software and archiving of all seeing data.

Apart from roof opening, the MASS-DIMM has been working in fully robotic
operation since March 2011. On all clear nights the system was operated by 
one of the SALT operators.

In summary, the current system is made up of a 25 cm telescope on an
equatorial pier at ground level, the entrance aperture of the telescope 
being at approximately 1 m from the ground. Targets are automatically 
selected to have a magnitude lower than 3 and a zenith angle less than 
45$^o$. The MASS data are processed by the Turbina 2.06 software , and 
the DIMM data are managed by the TimDIMM software, which also control the 
whole system. All of the data, raw and reprocessed, are stored within the SALT 
Science Archive.  

\subsubsection{Data processing}
 
Prior to using the TimDIMM setup on the DIMM channel of the MASS-DIMM,
the DIMM data were handled by the RoboDIMM software. A description of
the latest version is given in \citet{b19}, although we used a former
version, Robodimmnet 1.4. In this configuration, the DIMM frame
rate was set to 2 and 4 ms exposure time, allowing a correction of the
finite exposure time to the zero exposure time, as
described in \citet{b25}. The latest configuration uses the 
TimDIMM setup described in the previous section.
 
The MASS data are acquired by the Turbina software and reprocessed by
the Atmos algorithm \citep{b26}. The Atmos algorithm fits the measured
SI and DSI to atmospheric models, assuming Kolmogorov turbulence, in
order to restore the fixed and floating layers turbulence profile (\citet{b26} 5.1.2.10). 
In this study we will use results from the fixed layers method which is 
tuned to give the C$_N^2$ value at altitudes 0.5, 1, 2, 4, 8, and 16 km. 
Concerning the data acquisition, the PMTs have a 1 ms exposure time. Scintillation 
indices and statistical moments of the photon counts are calculated over a 
base-time of 1 sec, corresponding to hundred 1 ms samples. Those values 
are then averaged over a 1 min accumulation time, corresponding to 60 data point. After each accumulation time, atmospheric parameters are calculated from the SI, DSI and their respective
errors, and then stored in the output file \citep{b28}.
 
Previous studies of MASS and DIMM data quality indicate that data of low quality should be rejected \citep{b23,b27}. For DIMM data, the criteria are the number of frames per integration time (1 min) and the signal-to-noise ratio (SN) of the star images. For MASS data, we consider the flux in channel D, the error on the the flux measurement, the scintillation in channel A, and the $\chi^2$ of the restored profiles. A summary of the chosen threshold values is given in Table~\ref{tresh}. We chose the threshold levels in agreement with those indicated in \citet{b23}, except for the flux level in channel D. The flux treshold accounts for clouds and target lost, \citet{b23} indicates to choose a treshold of a minimum of 100 counts. However measurements with counts in chanel D as low as 50 are consistent with the rest of the data set from the same run. Hence, we choose to apply a treshold of 50 for the flux. The limit on the $\chi ^2$ value allows to keep only data for which the reconstruction of the TP is accurate, and the threshold on the SI value suppresses data taken under strong scintillation for which the MASS measurements are not accurate \citep{b28}. In order to be able to get a reliable value for the ground layer from differential DIMM and MASS measurements, we set the maximum value of $S_A^2$ to 0.7 as indicated in \citet{b23}. The percentage of data rejected for each thresholds never exceeds 5\%, and the parameters values obtained from the raw data are within $\pm$ 6\% of the ones obtained from the "cleaned" data. For DIMM data, a flux threshold is also applied to each aperture along with a threshold on the number of frames used to calculate the variance of the differential image motion.  This avoids biasing due to low signal-to-noise and helps to keep the statistics representative by maintaining a minimum number of samples per measurement.

\begin{table}
 \centering
  \caption{Data reprocessing thresholds}
  \begin{tabular}{@{}l||c|c@{}}
  \hline
        & Threshold parameter & Discrimination value\\
  \hline
  \hline
   MASS & Flux in D channel & $>50$ \\
        & Flux error & $<0.01$ \\
        & $\chi ^2$ & $<100$ \\
        & $S_A^2$ & $<0.7$ \\ 
  \hline
   DIMM & Number of rejected frames & $<50$ \\
        & S/N & $>5$ \\
\hline
\end{tabular}
\label{tresh}
\end{table}
\begin{table*}
  \tiny
  \caption{Observing periods with the Sutherland seeing monitoring instruments. The telescope diameter and DIMM operating and data processing softwares are given in parenthesis.}
  \begin{tabular}{@{}l||rr|rrrrrr|rr@{}}
  \hline
        & \multicolumn{2}{c}{{\bf SLODAR}} & \multicolumn{6}{c}{{\bf MASS-DIMM}} & \multicolumn{2}{c}{{\bf secondary DIMM}}\\
        & \multicolumn{2}{c}{{\bf (40 cm)}} & \multicolumn{6}{c}{{\bf (25 cm)}} & \multicolumn{2}{c}{{\bf (25 cm)}}\\
        \cline{4-9}\\
        &  &  & \multicolumn{2}{c}{MASS} & \multicolumn{2}{c}{DIMM (RoboDIMM)} & \multicolumn{2}{c}{DIMM (TimDIMM)} & \multicolumn{2}{c}{(TimDIMM)}\\        
   Dates & $\#$ of nights & $\#$ of data & $\#$ of nights & $\#$ of data & $\#$ of nights & $\#$ of data & $\#$ of nights & $\#$ of data & $\#$ of nights & $\#$ of data\\
 \hline
 February 2010 & 8  & 1355 & -- &   -- & -- & -- & -- & -- & -- & -- \\
 March 2010    & 16 & 2534 & 8 & 2108 & 8 & 2260 & -- & -- & -- & -- \\
 April 2010    & 4  & 646 & 20 & 7094 & 20 & 7817 & -- & -- & -- & -- \\
 May 2010      & -- & -- &  4 & 1180 &  4 & 1230 & -- & -- & -- & -- \\
 June 2010     & -- & -- &  1 &   -- &  1 &   -- & -- & -- & -- & -- \\
 July 2010     & -- & -- & 10 & 3983 &  7 & 2199 & -- & -- &  3 & 1272 \\
 August 2010   & -- & -- &  4 &  980 & -- &   -- & -- & -- &  8 & 3990 \\
 September 2010 & -- & -- & -- &  -- & -- &   -- & -- & -- & 18 & 8436 \\
 October 2010  & -- & -- & -- &   -- & -- &   -- & -- & -- &  3 &   -- \\
 November 2010 & -- & -- &  1 &  239 & -- &   -- & -- & -- & 14 & 7980 \\
 December 2010 & -- & -- &  2 &  614 & -- &   -- & -- & -- & 14 & 2909 \\
 January 2011  & -- & -- & 16 & 4889 & -- &   -- & -- & -- & 21 & 7500 \\
 February 2011 & -- & -- & 14 & 4277 & -- &   -- & 14 & 7045 & -- & -- \\
 March 2011    & -- & -- & 11 & 3818 & -- &   -- & 11 & 6001 & -- & -- \\
\hline
\end{tabular}
\label{obs_T}
\end{table*}
\subsection{Observing periods}

This first year of seeing monitoring was conducted with several
instruments that were operating at different periods, locations, and with 
different telescopes. Table~\ref{obs_T} gives an overview of the different operation
schedules for each instrument. Simultaneous operation of the SLODAR
and MASS-DIMM were carried out over a period of 10 nights in March
2010. Also, from August 2010 to early January 2011, more DIMM than
MASS data have been taken. This is due to the fact that the TimDIMM
was set to be remotely controlled, while the MASS-DIMM still required
manual operation. Only the TimDIMM was operated for most of the nights
through this period. Data from June and October 2010 were included
into the statistics for July and November respectively. Only few
nights were observed during those months, and these nights were at the
beginning or the end of the month.

\section{Data analysis and consistancy}

\subsection{Data analysis method}\label{sec:analysis}

For the purpose of data analysis, we define the free atmo- sphere (FA) as all layers above 1 km. We re-derive the FA seeing from the CN2 profile measured by the MASS consid- ering only the layers at and above 1 km and excluding the measurements at 500 m from MASS, which are known to be inaccurate \citep{b39}. However, the Atmos soft- ware, which computes the MASS atmospheric profile, uses triangular weighting functions to define the altitudes of the layers \citep[Fig 2. in ][]{b29}. Due to the layer boundaries not being sharply defined, the reported CN2 value at a given altitude will include contributions from a range of heights. For example, the 1 km layer includes contributions from turbulence located between 500 m and 2 km. As such, any turbulence measured by MASS at a specific altitude is a weighted average of turbulence measured over the altitude range of the triangular weighting function. For simplicity in the rest of the paper, we will use 1 km as our boundary for the FA, but keeping in mind that this threshold is only approximate.

Defining the GL-FA boundary at 1 km requires a re-binning (in terms of
altitude and time) of the data from the MASS-DIMM.  The total
integrated turbulence profile is given by:
\begin{equation}
  J = \int_{0}^{\infty} C_N^2(h) dh.
\end{equation}
The integrated turbulence profile can be given for different parts of
the atmosphere by integrating $C_N^2 (h)$ over different heights.  For
example, $J^{500}$ is given by integrating between 250 m and 1 km
and using the triangular weighting function. $J^{MASS}$ is the sum of all layer contributions,
$J^{500} + J^{1} + J^{2} + J^{4} + J^{8} + J^{16}$. The model gives a discrete representation of the turbulence which is in reality continuously distributed. If the real turbulence is at an altitude located in between two of the predefined layers altitude, it will be redistributed between those 2 adjacent layers through the triangular weighting functions (see fig.~6 and 7 in \citet{b28}). It is shown in section 6.2 of \citet{b28} that the redistribution induces errors on the individual
layers contribution, reaching 20\% in the worst cases for the lower 500 m and 1 km layers,
but is typically between 5 to 10\% for the overall contribution, $J^{MASS}$. 
The seeing in arcseconds for a given layer is just obtained through the
following relation: $\epsilon = J^{3/5} \times 5.307\lambda^{-1/5} \times 206265$, where
$\lambda$ is in $m$ and J is in $m^{1/3}$.

To determine the free atmosphere seeing, we must subtract the turbulence
contribution of the 500 m layer from the overall MASS turbulence, then calculate
the corresponding seeing. Using MASS profiles, the FA seeing ($\epsilon ^{FA}$) is given by:

\begin{equation}
\epsilon ^{FA} = [J^{MASS} - J^{500}] ^{3/5} \times 5.307 \lambda ^{-1/5} \times 206265.
\end{equation}

The GL seeing is then calculated as follow, after the 10 min binning:
\begin{equation}
\epsilon ^{GL} = [(\epsilon ^{DIMM}) ^{5/3} - (\epsilon ^{FA}) ^{5/3}] ^{3/5}.
\end{equation}

In terms of qualitative analysis we will define best seeing conditions
to be the first quartile of the seeing distribution, median conditions
corresponds to the 25\% around the median value, and bad conditions
will be the fourth quartile of the seeing distribution.

\subsection{Data consistency}

\subsubsection{MASS vs. DIMM}
\begin{figure}
\centering
\includegraphics[scale=0.4]{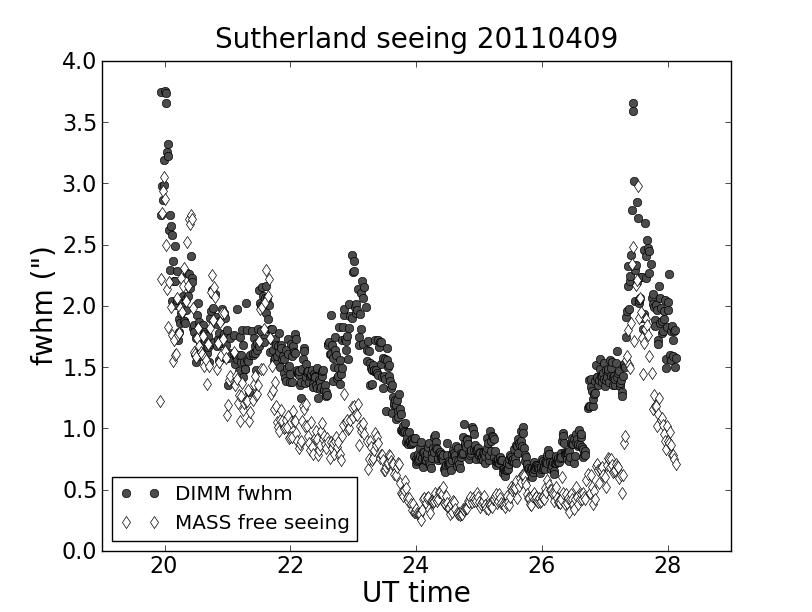} 
\caption{An example of the nightly variation of the overall seeing
  measured by DIMM (filled circle) and the seeing measured by MASS from layers
  above 500 m (open diamond).}
 \label{MD_night}
\end{figure}

Despite using two different technologies for measuring the seeing, the
DIMM and the MASS show consistent behaviour throughout a typical
night. An example of this is given in Fig. ~\ref{MD_night}. As expected,
$\epsilon ^{DIMM} \geq \epsilon ^{MASS}$, since the DIMM is measuring
the overall seeing while the MASS is only measuring the FA seeing.
However, on nights when the FA is dominant and one would expect
$\epsilon ^{MASS} \approx \epsilon ^{DIMM}$, $\epsilon ^{MASS}$ can be
$>$ $\epsilon ^{DIMM}$ due to the overestimation of the contribution
of the lower layers to the MASS seeing \citep{b39}.

\subsubsection{DIMM vs. SLODAR}

\begin{figure}
\centering
\includegraphics[scale=0.4]{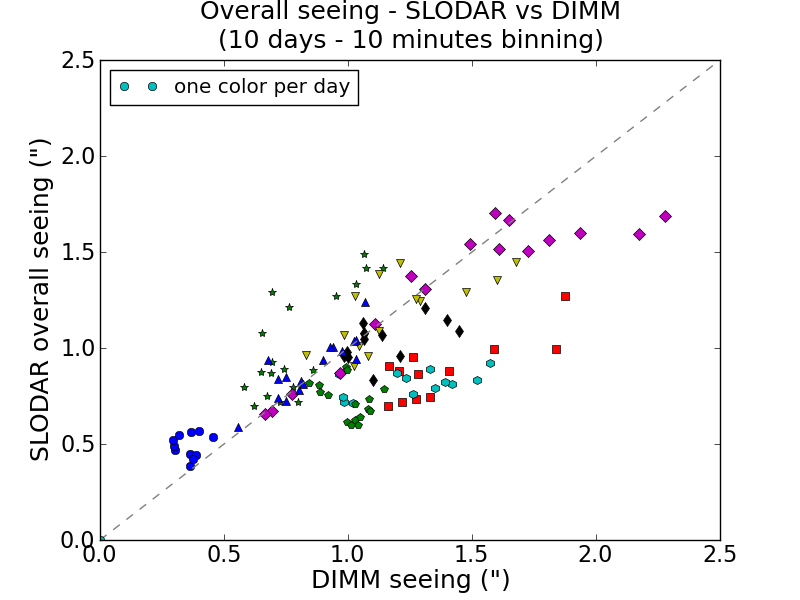}  
\caption{Comparison of the overall seeing value as measured by the
  DIMM and the SLODAR. Different data points and colors indicate
  different observing nights.}
 \label{SL_vs_D}
\end{figure}

As confirmation of the reliability of the seeing monitors, we compared
seeing values as measured by the DIMM and the SLODAR. In order to do
so, only data from overlapping observing periods have been considered.
In addition, data have been binned and averaged over 10 minute periods
to account for non-simultaneity of the measurements. The overall
correlation between DIMM and SLODAR seeing values can be seen in
Fig.~\ref{SL_vs_D}. Each night is represented by a different data point and color. Despite a high dispersion on certain nights, the DIMM and SLODAR measurements
are reasonably well correlated: the Pearson correlation coefficient, which measures 
linear relationships, is 0.69. A possible explanation
for the differences between the instruments seen on any given night
may be due to the different optical paths observed by each instrument
as they were separated by 200 m on the plateau and were not always
targeting the same star. In addition, the tendency of DIMM to give higher 
values can be explained by dome/tube seeing which, as mentioned in \S 3.1.4, 
has been removed from SLODAR measurements while it could still be affecting DIMM ones.

\section{Results}
\begin{figure*}
  \centering 
   \includegraphics[scale=0.4]{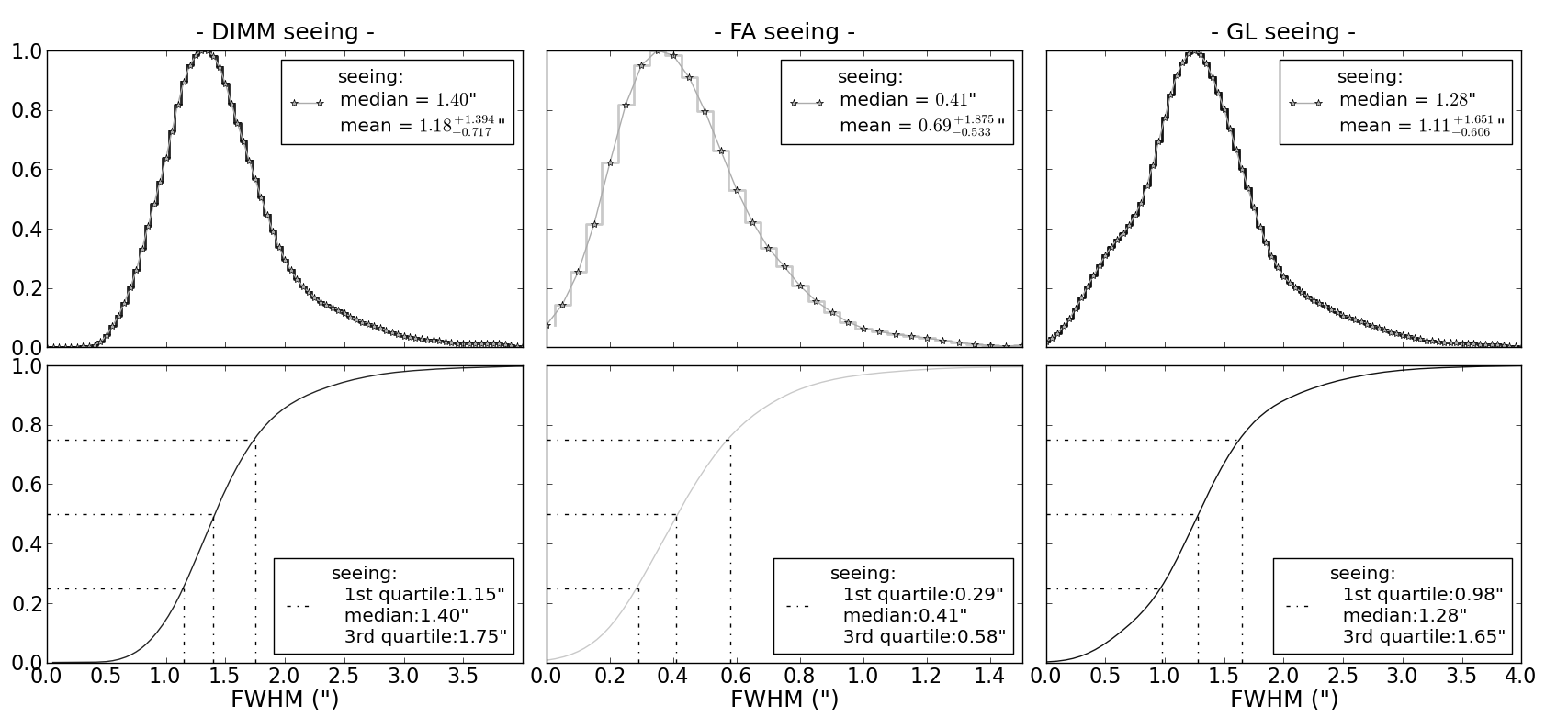}
\caption{First year statistics of the seeing at the Sutherland site. Data
  from MASS-DIMM measurements, with the division between GL and FA defined
  to be at 1 km above ground level. All data have been averaged over 10
  min. Left: Overall seeing as measured from DIMM. Center: Free-atmosphere
  seeing from MASS. Right: Ground-layer seeing derived from the difference
  between DIMM and MASS (see \S \ref{sec:analysis}).}
 \label{see_stat}
\end{figure*}
\subsection{General statistics of the seeing}
\label{sec:seeing}

We used all of the DIMM data obtained with both MASS-DIMM (until July
2010) and TimDIMM (afterwards) with no temporal averaging to compute
the median seeing for the site.  The statistics for this present study
are made of 53938 data points taken over a period of one year, from
March 2010 to March 2011 and covering 146 nights. The median seeing
value is 1.32".  

Using only the measurements for which both MASS and DIMM data are
available, we looked at the contribution from the FA and the GL to the
overall seeing.  Distributions of the seeing values associated with
each of these three components are shown in Fig.~\ref{see_stat}.  To compare
the MASS and DIMM measurements, we have averaged the data over 10
minute periods as explained in \S \ref{sec:analysis}. Those results
consist of 2564 data points observed over 91 days throughout the
year, from March 2010 to March 2011.  For this data set, the median
overall seeing was 1.4" (Fig.~\ref{see_stat}, left), the median FA seeing was
0.41" (Fig.~\ref{see_stat}, middle) and the median GL seeing was 1.28"
(Fig.~\ref{see_stat}, right).
\subsection{Atmospheric turbulence profiles}

\subsubsection{Turbulence profiles from MASS-DIMM}

%
\begin{figure*}
\centering
\subfigure[]{
\includegraphics[scale=0.35]{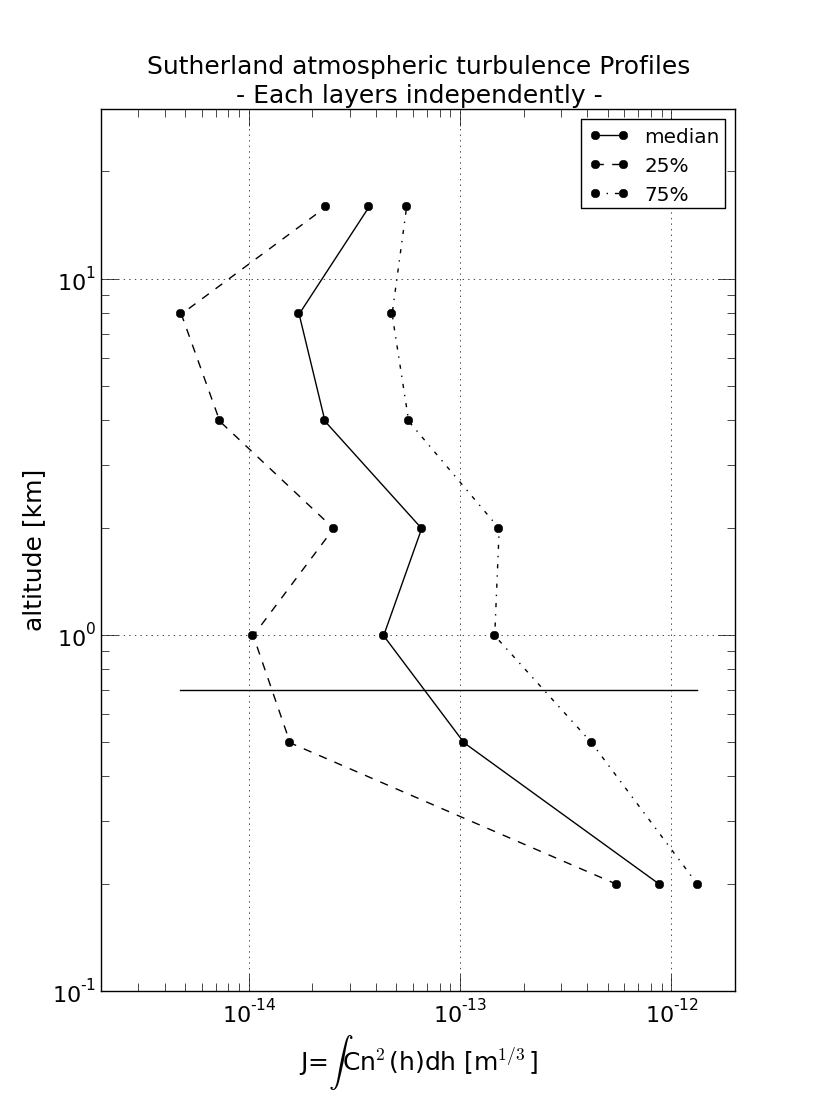}   
}
\subfigure[]{
\includegraphics[scale=0.35]{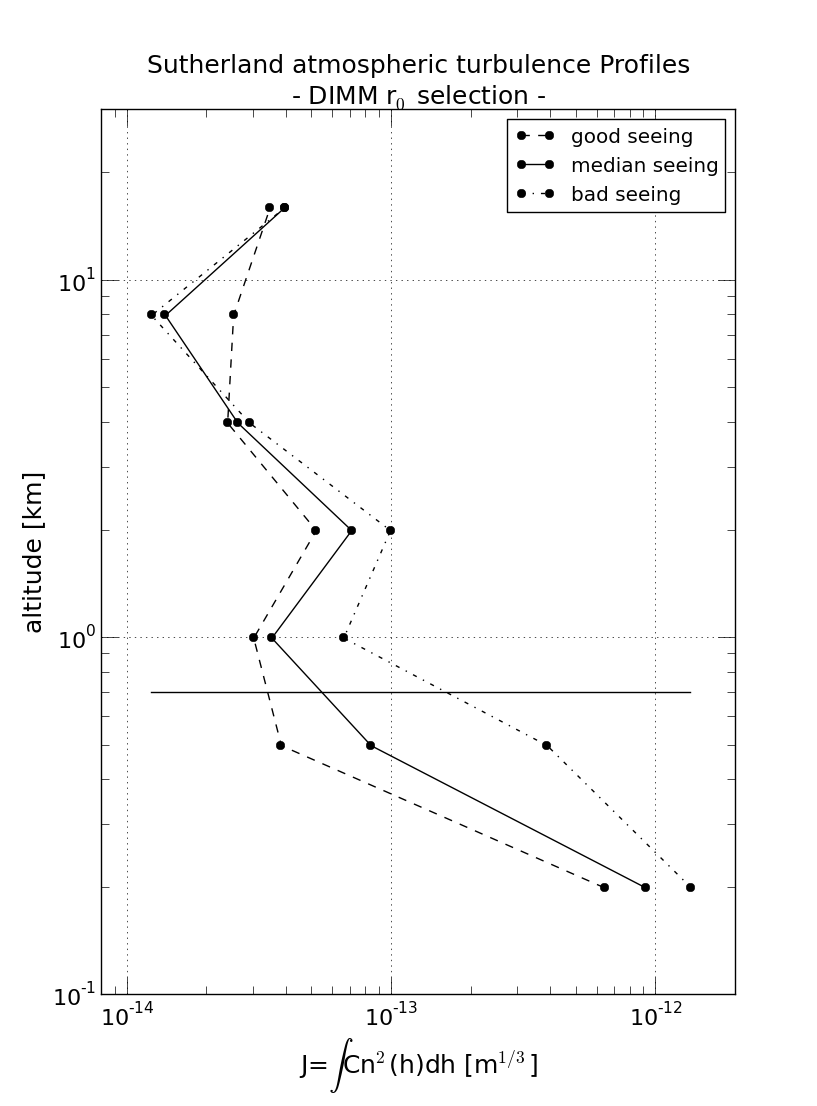}
}
\caption{Turbulence profiles from MASS-DIMM: (a) gives the median (solid line), fisrt (dashed line) and third quartile (dotted line) value of the turbulence strength for each layer independantly, (b) represents the typical profiles under "median" (solid line), "good" (dashed line) and "bad" (doted line) seeing conditions. "Good", "median" and "bad" are defined as the 20-30\%, 45-55\% and 70-80\% ranges of the respective cumulative distribution of $\epsilon _0$.}
\label{MD_profile}
\end{figure*}
\begin{table}
 \centering
 \caption{Contribution of the different layers for the median profile (solid line in Fig.~\ref{MD_profile}-b)}
  \begin{tabular}{@{}c||c|c|c|c@{}}
    & layer & \multicolumn{2}{c}{layer contribution} & seeing \\ 
    & altitude & \multicolumn{2}{c}{to the turbulence} & \\
   \hline
    & 16 km & 3.3\% & & \\
    & 8 km & 1.2\% & & \\
    FA & 4 km & 2.2\% & 15.7\% & 0.41" \\
    & 2 km & 6\% & & \\
    & 1 km & 3\% & & \\
   \hline
    & 500 m & 7\% & & \\
    GL & & & 84.3\% & 1.28" \\
    & 200 m & 77.3\% & &  \\
    \hline
    \hline
    Overall & & & & 1.4" \\
    \end{tabular}
    \label{profile}
\end{table}
In addition to the integrated parameters, the MASS-DIMM gives low
resolution profiles of the turbulence. Fig. ~\ref{MD_profile} shows the typical turbulence profiles at the Sutherland site. Fig.~\ref{MD_profile} (a) gives the median, first and third quartile value of the cumulative distribution of each layer independently. In Fig.~\ref{MD_profile} (b) we represent the profiles as selected by their associated $r_0$ value. "Good" correspond to profiles for which the associated $r_0$ is within the 20-30\% interval of its cumulative distribution. "Median" correspond to the 45-55\% interval and "bad" is the 70-80\% interval. The choice of interval was based on similar studies in the litterature \citep{b4,b53}. The contribution of each layers in terms
of $J$ value is represented for layers at different altitude. The
FA layers are those given by the 1,2,4,8 and 16 km by the MASS. The GL has been 
divided into two layers: a 500 m layer also given by MASS and a 200m layer that includes all lower layers. The 200 m altitude was determined
by taking the average altitude ($h_{eq}$) of the turbulence weighted by the
turbulence strength at each altitude from SLODAR ground layer data, 
$h_{eq} = \frac{\sum h_i . J_i}{\sum J_i}$, which gives a value of 227 m.

Fig.~\ref{MD_profile} (a) only gives the median profile and the range of turbulence strength for each layer with the first and 3rd quartile values. While less representative of realistic profiles, due to the fact that it is unlikely to be in a configuration where all layers will be in their best condition or worse condition simultaneously, this representation is relevant for site comparison purpose.

Fig.~\ref{MD_profile} (b) is a more relevant plot in terms of site characterization as it gives the typical profiles corresponding to "good", "bad", and "median" seeing conditions. One can see that, as the seeing degrades, the 200 m, 500 m, 1, and 2 km layers become more turbulent. While having a weaker contribution to the overall turbulence, the higher layers from 4 km and above remain, overall, fairly similar under all seeing conditions. However, the 8 km layer appears to be more turbulent under good seeing than for median and bad seeing. The overall contribution of the GL increases as the seeing degrades, being 80\% under good seeing conditions (dashed line), 84\% under median seeing (solid line) and nearly 88\% under bad seeing (dotted line). Also, while the distribution between the two GL layers at 200 and 500 m does not change much between good and median seeing, it varies significantly under bad seeing conditions with 68\% of the overall turbulence in the 200 m layer and 20\% in the 500 m as compare to 77\% and 7\%, respectively, under median seeing. As the seeing degrades the 500 m layer contributes more to the GL turbulence. We will discuss this point further in the following section with the results of the GL profiles from SLODAR.

Overlall, based on the comparison of three typical profiles under good, median and bad seeing, one can conclude that the seeing value is mainly driven by the ground layer turbulence and to a weaker extent by the wind shear layer in the free atmosphere
located around 2 km. The contribution from the different layers under
median seeing conditions is summarized in Table~\ref{profile}.
\subsubsection{Ground layer turbulence profiles from SLODAR}
%
%
%
\begin{figure*}
\centering
\subfigure[]{
\includegraphics[scale=0.30]{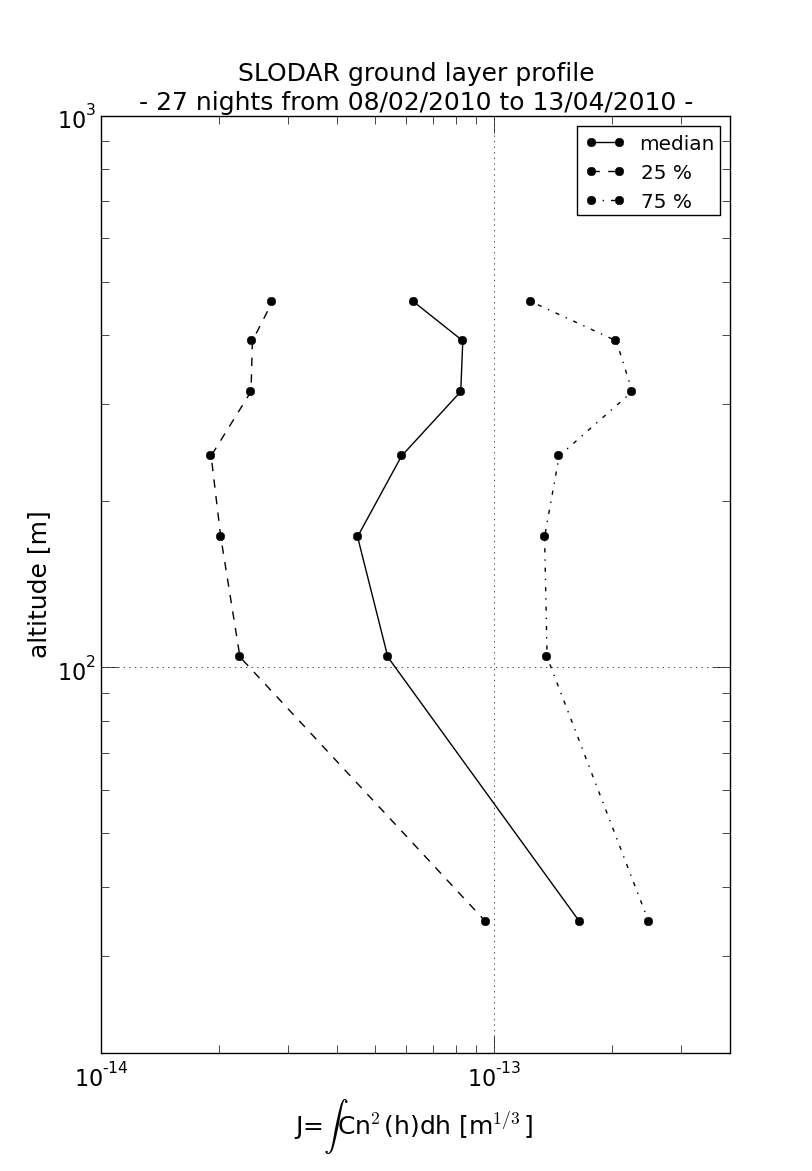}
}
\subfigure[]{
\includegraphics[scale=0.30]{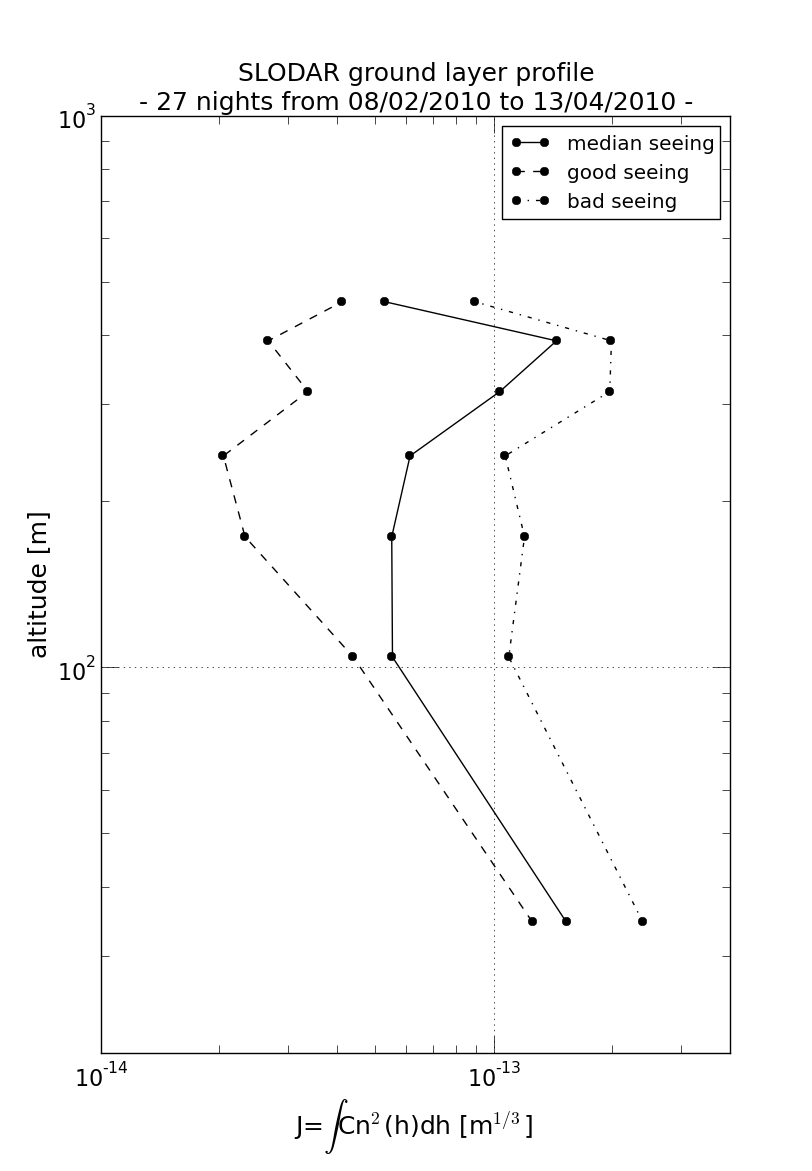}
}
\caption{GL turbulence profiles from SLODAR, based on 27 nights of data between February and April 2010. (a) gives the median (solid line), fisrt (dashed line) and third quartile (dotted line) value of the turbulence strength for each layer independantly. (b) are the typical profiles under "median" (solid line), "good" (dashed line) and "bad" (doted line) GL seeing conditions. "median", "good" and "bad" are defined as the 45-55\%, 20-30\% and 70-80\% ranges of the cumulative distribution of $\epsilon_0 ^{GL}$.}
\label{SL_profile}
\end{figure*}
\begin{table}
 \centering
 \caption{Contribution of the different layers for the GL median profile from SLODAR (solid line in Fig. \ref{SL_profile}-b)}
 \label{tabgl}
  \begin{tabular}{@{}c|c@{}}
    layer altitude & layer contribution \\ 
   \hline
    35 m & 24.5\%\\
    105 m & 8.8\%\\
    173 m & 8.8\%\\
    243 m & 9.8\%\\
    318 m & 16.6\%\\
    393 m & 23.1\%\\
    462 m & 8.4\%\\
    \end{tabular}
\end{table}
From the 27 nights of SLODAR measurements, we extracted the typical
GL profiles. SLODAR profiles provide the turbulence strength for 7 layers whose altitudes vary slightly with both the separation and zenith angle of the target double star. For our typical profiles we determined each layer altitude 
by taking a weighted average, $h^{layer} = \frac{\sum h^{layer}_i . 
J^{layer}_i}{\sum J^{layer}_i}$, over all observations. Similarly to the profiles from MASS-DIMM we have represented the median, first and third quartiles for each layers independently on Fig.~\ref{SL_profile} (a), similar as to the measurements for Mauna Kea by \citet{b5}. Fig.~\ref{SL_profile} (b) shows the profiles correponding to "good", "median" and "bad" seeing condition with the same intervals as used for the MASS-DIMM profiles, but here using only the GL seeing value rather than the overall seeing. This method is similar to the measurements of Mount Graham by \citet{b48}.

Looking at the contribution from each layer for the profiles associated to the median seeing conditions (solid line in Fig.~\ref{SL_profile} (b) and summarized in Table ~\ref{tabgl}),  the first layer at 35 m contributes nearly 25\% of the GL seeing while approximately 50\% is contributed by layers above 300 m. On the other hand, under good seeing conditions nearly 40\% of the turbulence is within the first 30 m while the upper layers are much weaker, particularly the 400 m that is now less than 10\% of the turbulence. Under bad seeing the contribution of the lower 30 m goes down to 20\% and the 300 to 500 m contribute around 45\%, while the contribution from the intermediate layers, 100 to 300 m, increases to 30\%. The degradation in the GL seeing corresponds to the increasing turbulence in the upper layers. This is in good agreement with our earlier observations from MASS-DIMM profiles which show an increase of the 500 m contribution as the seeing degrades.

In Section 6.2, we compare the profile of the GL at Sutherland with that measured at other sites. 
\begin{figure*}
\centering
\subfigure[]{
  \includegraphics[scale=0.4]{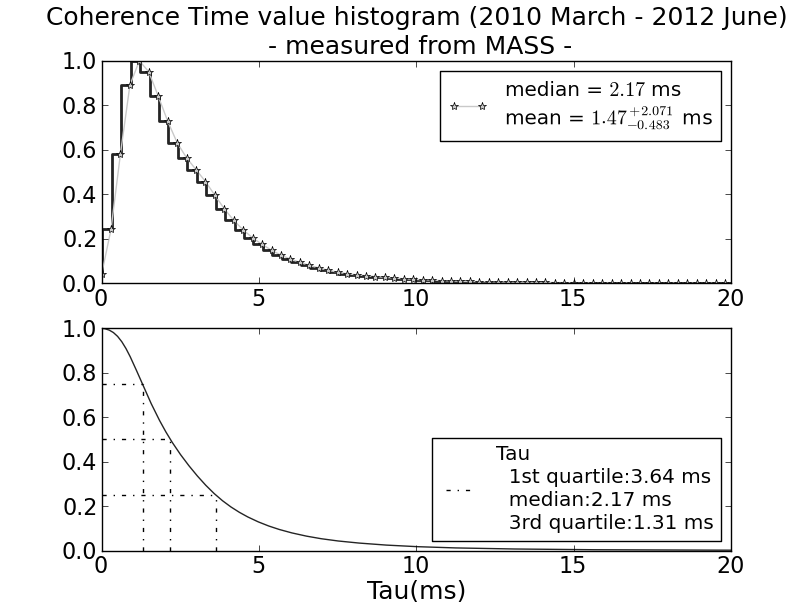}
}
\subfigure[]{
  \includegraphics[scale=0.4]{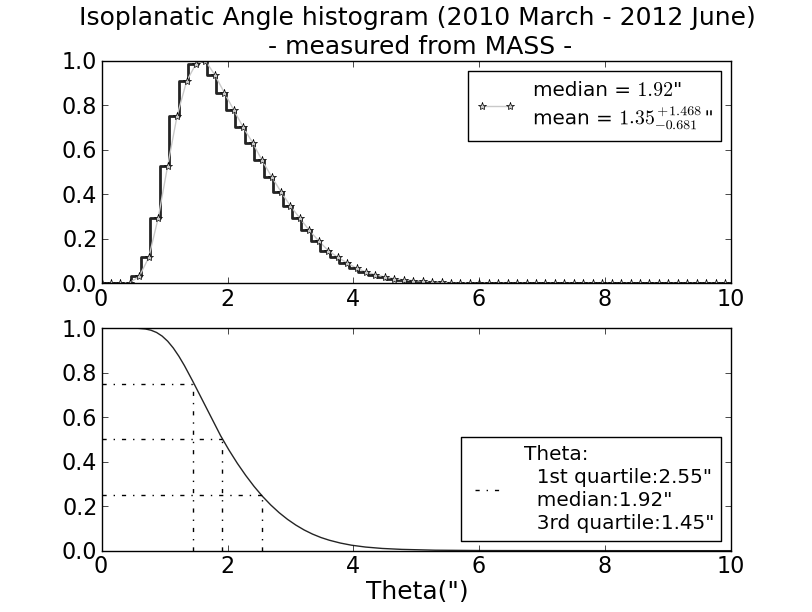}
}
 \caption{First year statistics of the seeing at the Sutherland site. MASS data. (a) Coherence time of the turbulence. (b) Isoplanatic angle.}
 \label{T_iso_stat}
\end{figure*}
\subsection{Isoplanatic angle and coherence time}
\label{sec:iso_time}
\begin{table}
 \centering
 \caption{Coherence time at Sutherland. Value given by the MASS instrument ($\tau _0 ^{MASS}$) and corrected value ($\tau _0 ^{corr}$). The correction uses $C=1.73$ and the GL contribution via $J_{GL}$ in equation (4). The wind speed was obtain from the weather mast data 30 m probe ($5.1 m.s^{-1}$).}
 \label{Tau0_tab}
  \begin{tabular}{@{}l||c|c|c|c|c@{}}
     & median & 25\% & 75\% & \multicolumn{2}{c}{{\large $\epsilon _0$}} \\ 
     & & & & best 5\% & best 20\% \\ 
   \hline
    $\tau _0 ^{MASS}$ [ms] & 2.17 & 3.64 & 1.31 & 8.87 & 4.33 \\
    \\
    $J_{GL}$ [$10^{-13}$ $m^{1/3}$] & 8.99 & 5.88 & 14.02 & 0.32 & 4.92 \\
    (DIMM-MASS) &  &  &  &  & \\
    \\
    $\tau _0 ^{corr}$ [ms] & 2.85 & 4.28 & 1.86 & 13.65 & 4.93 \\
    (MASS$_{corr}+$GL) &  &  &  &  & \\
    \hline
    \end{tabular}
\end{table}
In addition to the seeing value, the MASS also provides a measurement
of both the coherence time ($\tau _0$) and the isoplanatic angle
($\theta _0$). Statistics of those two parameters are presented in
fig. ~\ref{T_iso_stat}.\\

The median value of the coherence time ($\tau _0$) at Sutherland, as given from the MASS instrument, is 2.17 ms, the first and third quartile are respectively 3.64 and 1.31 ms. However it has been shown that the MASS measurement underestimates the coherence time \citep{b49}, but a correction can be applied. We will use equation (5) given in \citet{b49}:
\begin{equation}
 \tau _0 ^{-5/3} = (C . \tau _{MASS})^{-5/3} + 0.057^{-5/3} \lambda _0 ^{-2} V_{GL}^{5/3} J_{GL},
\end{equation}
where $\tau _{MASS}$ is the value given by the MASS, $\lambda _0$ is the wavelength (500 nm), $V_{GL}$ is the ground layer wind speed obtained from the 30 m high probe on the  weather mast ($5.1 m.s^{-1}$ ) and $J_{GL}$ is the strength of the turbulence in the ground layer, derived from $'DIMM-MASS'$. $C$ is a correction coefficient, whose value is determined empirically with fairly low accuracy. Tokovinin found a value of 1.27, however he also mentioned a value of 1.7 and 1.73 found by \citet{b51} and \citet{b50} as well as values varying between 2 and 2.5 found with comparative data at Cerro Tololo and Paranal. For the present study we will use $C=1.73$, given in \citet{b50}, as it is the most commonly used, and in particular for the TMT site testing. This gives us a consistent comparison between the Sutherland site and other sites around the world, keeping in mind that the accuracy of the method is $\pm$ 20\%.

Results before and after applying the correction are presented in Table ~\ref{Tau0_tab}. After correction, the median value, first and third quartiles of the coherence time at Sutherland are respectively 2.85, 4.28 and 1.86 ms. We also looked at the coherence time value under good seeing conditions, for the {\large $\epsilon _0$} best 5\% and 20\% values. We found, respectively, a 13.65 ms and 4.93 ms coherence time. 
Comparison values from 12 other sites are given in Table~\ref{othersites}.

The isoplanatic angle median value at Sutherland is 1.92". The values for 12 comparison sites are listed in Table~\ref{othersites}. The first and third quartile values are respectively 
2.55" and 1.45", which gives a range of 1.1" between the best 25\% and 75\%. 
This is similar to all sites apart from Mauna Kea and Mount Graham spanning over a 
2" range with best 25\% of respectively 3.61" and 3.6" \citep{b32,b47}.
\subsection{Correlation of seeing with the weather conditions}
\begin{figure*}
\centering
\subfigure[]{
\includegraphics[scale=0.16]{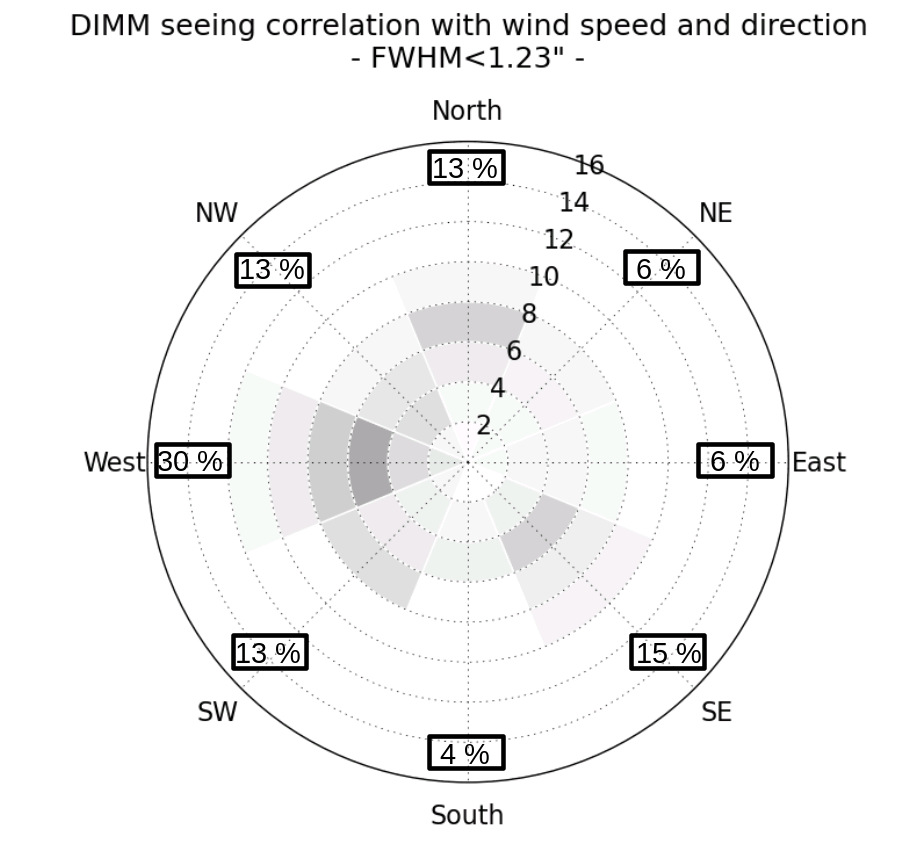}   
}
\subfigure[]{
\includegraphics[scale=0.16]{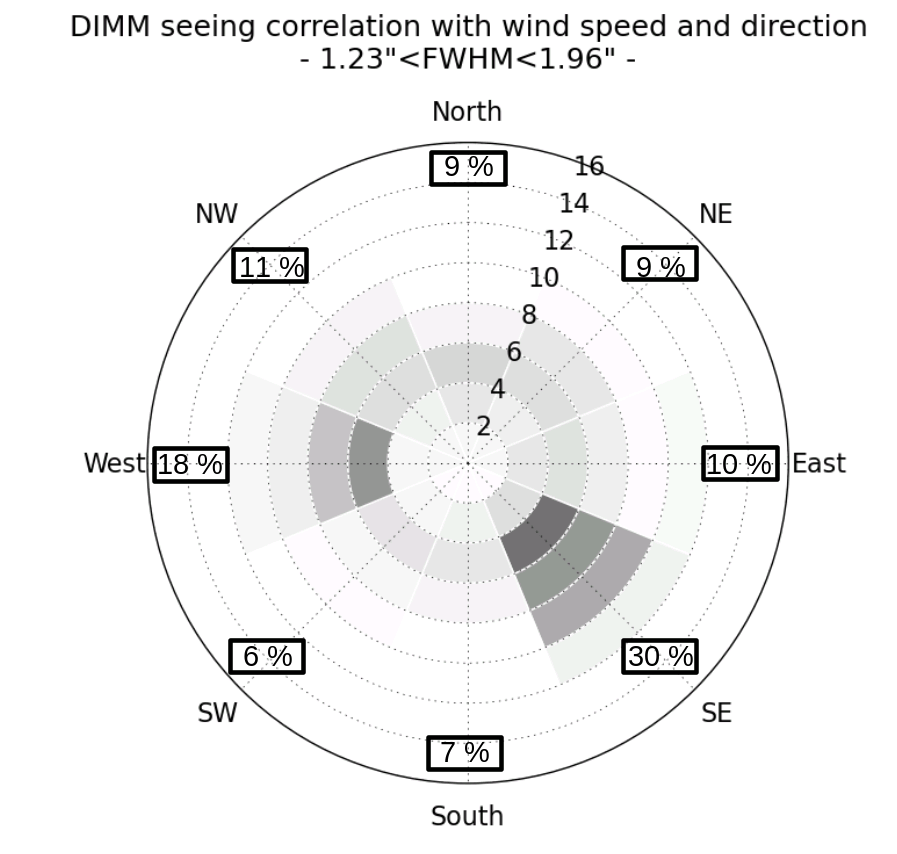}
}
\subfigure[]{
\includegraphics[scale=0.16]{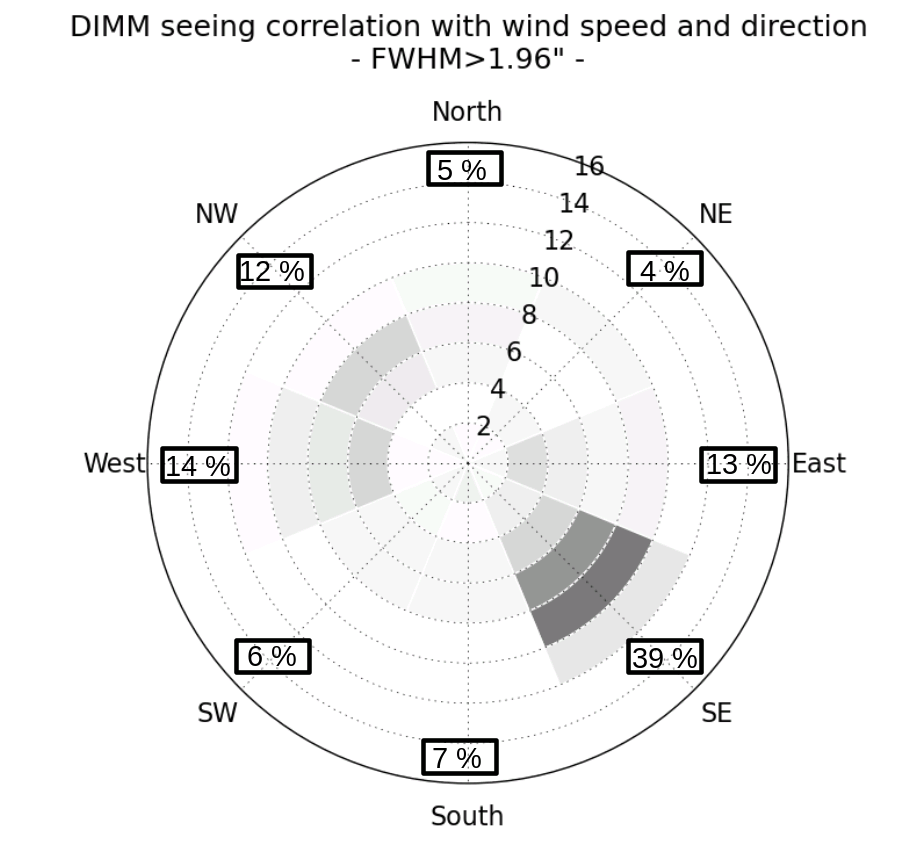}
}
\subfigure{
\includegraphics[width=8mm,height=50mm]{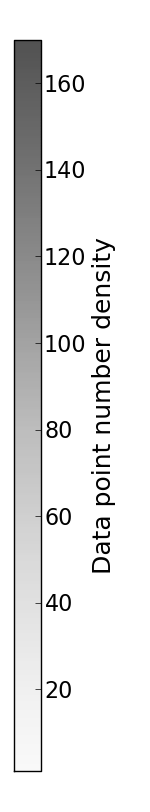}
}
\caption{Correlation of seeing with wind direction and speed. In all three 
  figures, the angle corresponds to the wind direction, the radius
  gives the wind speed in m/s, and the color gradient gives the
  density of data points. a) Good seeing conditions: $\epsilon_0 \le
  1.23"$, corresponding to the first quartile. b) Median seeing
  conditions: $1.23" \le \epsilon_0 \le 1.96"$. c) bad seeing
  conditions: $\epsilon_0 \ge 1.96"$, corresponding to the third
  quartile. The percentages indicate the contribution of each wind directions in 
  each wind rose.}
\label{wind_dir}
\end{figure*}
\begin{figure*}
\centering
\includegraphics[scale=0.4]{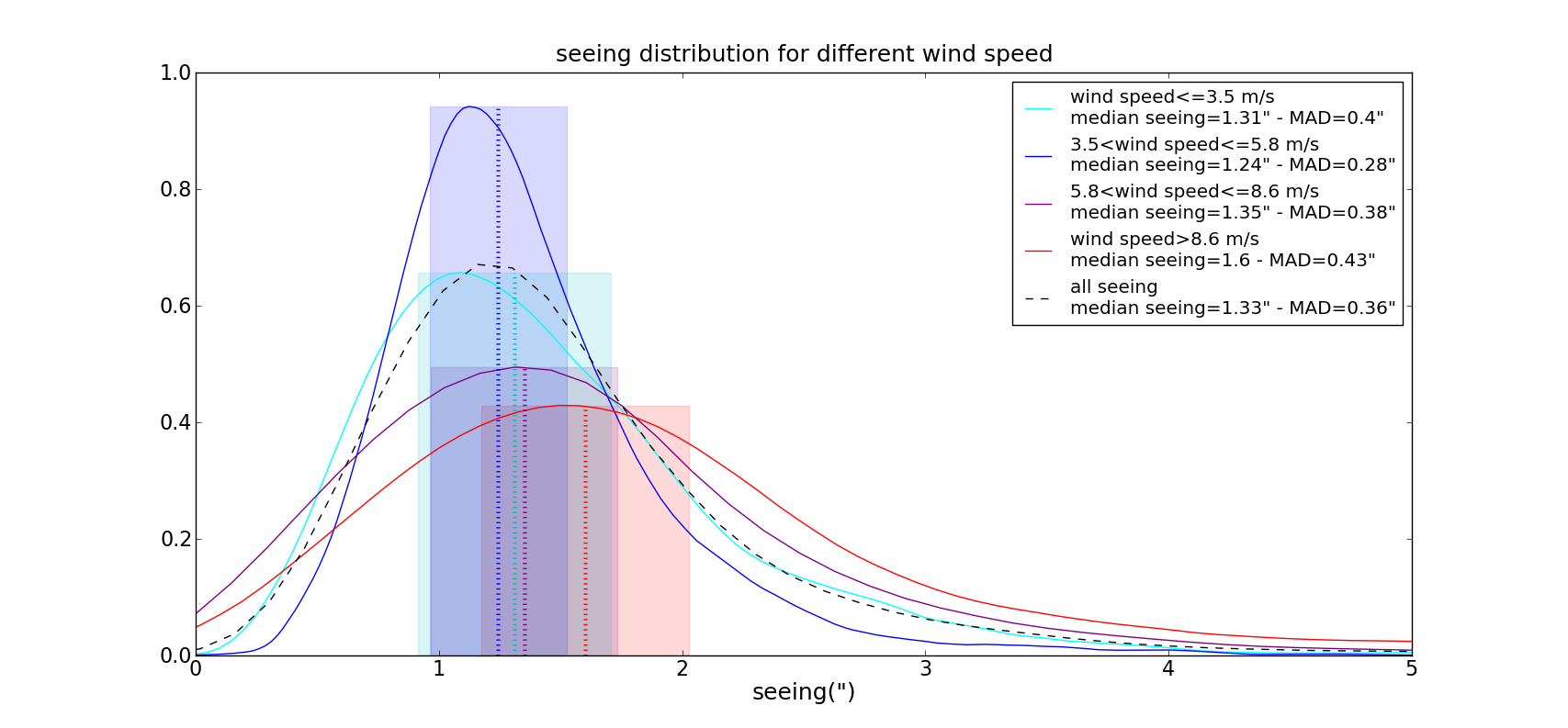}
\caption{Relation between wind speed and seeing values. Dark blue:
  seeing distribution for lowest wind speed ($\le$ 3.5 m/s). Light
  blue: seeing distribution for wind speed between 3.5 m/s and 5.8
  m/s. Purple: seeing distribution for wind speed between 5.8 m/s and
  8.6 m/s. Red: seeing distribution for highest wind speed ($\ge$8.6
  m/s). The dashed line is plotted as an indicator and is the seeing
  distribution for all data. For all curves, the dotted line shows the
  median value and the shaded area corresponds to the median average
  deviation (MAD).}
\label{W_speed}
\end{figure*}
To see how the weather conditions influence the seeing, we compare our
seeing measurements with data from the SALT weather station, which
provided wind speed and direction at 10 and 30 m above ground,
relative humidity level, and temperatures at 2,5,10,15,20,25,30 m
above ground.  Previous studies \citep{b12} showed that the wind was
the main weather component influencing the seeing conditions. In
examining the current weather data, we do find a strong correlation
with wind direction, a weak correlation with wind speed, and we find
no correlation with the temperature or relative humidity.

\subsubsection{Influence of the wind direction} \label{res:wind}

Fig.~\ref{wind_dir} presents the correlation of the seeing with wind speed
and direction. We have plotted different cases corresponding to good
seeing conditions (a), median seeing conditions (b), and bad seeing
conditions (c). On the wind roses the angles indicate the wind
direction and the radii the wind speed in m/s.  After splitting the
data according to their seeing value in the 3 different seeing
conditions, we binned them within wind speed range of 2 m/s and wind
directions range of 45$^o$. The color of each bin indicates the
number of data within each bin, darker color indicating a higher
number density of events. In addition, for each wind rose, we indicate the overall contribution of each wind direction in percentages of all data within each seeing range. Looking at Fig.~\ref{wind_dir} (a), one can see that good seeing occurred predominantly under westerly winds, while Fig.~\ref{wind_dir} (c) shows that bad seeing happened when winds were coming from the South-East. Apart from the North-West and North-East directions, having a fairly similar contribution under all seeing conditions, we could split the roses in two: the South-West to North half and the North-East to South half. If we look at the South-West to North half, the contribution from those wind directions gets weaker as the seeing degrades. On the other hand the North-East to South half direction contribute more and more with degrading seeing. It is clear here that as the wind turns West to South-East, the 2 predominant wind directions, the seeing degrades. Moreover, in the case of South-Easterly winds one can notice, by comparing the second and third roses, that the seeing tend to get worse with higher wind speed. However as we will see in the next section this is a behavior that is only visible in the case of South-Easterly winds.

\subsubsection{Influence of the wind speed}

The median seeing for 4 different wind
speed ranges, corresponding to $<$3.5 m/s, 3.5 to 5.8 m/s, 5.8 to 8.6
m/s, and $>$8.6 m/s are presented in Fig. \ref{W_speed}.  The wind speed
does not appear to have a very significant influence on the seeing,
except at the highest wind speeds. However, the shift towards worse
seeing for high wind speeds is not very pronounced, and may be
due to telescope shake or higher dome seeing rather than higher 
atmospheric turbulence.

One could also suspect a wind speed bias due to the fact that the
seeing monitors cannot operate at wind speed higher than 16 m/s. 
However winds above this speed were only
recorded 2\% of the time over the last three years.  Moreover, the
seeing value is only relevant within the conditions for which the
telescopes are operational. Since the operation limit at Sutherland 
for the small telescopes and for normal operations of SALT is a wind 
speed limit of 16 m/s, we can neglect the effect of wind bias.

\section{Discussion}
\subsection{Degradation in the Site Conditions}

The value for the median seeing reported here is worse than those that
were reported in earlier studies.  Due to the methodology used in
\citet{b41}, we do not have a reliable way to compare their results with our
study as discussed in \S \ref{sec:site}. Concerning the site testing
results from \citet{b12}, a more reliable comparison is possible as
those measurements were made using the same type of
instrumentation. This previous campaign reported a median seeing of
0.92", which is much lower than the 1.32" obtained from our entire set
of DIMM data. That can be explained, to some extent, by the longer
exposure time used in the previous study.

The bias in DIMM measurements due to long exposure times has been
previously investigated by \citet{b25}.  We can repeat that analysis
for the specific observing setups used for DIMM measurements in
Sutherland studies by stacking our exposures of 3.3 ms taken with
TimDIMM.  We estimate the bias in the DIMM measurements from
\citet{b12} by replicating their 10 ms exposure time. The variance of
the differential motion in the stacked image with a 9.9 ms exposure
time was 0.62 times the variance of the differential motion in the
individual 3.3 ms exposures. Applying this correction to the 0.92"
seeing reported by \citet{b12} results in a seeing value of 1.25" for
a 3.3 ms exposure time, which is not significantly different to our
measured value of 1.32". Although this result is similar to the
analytic model of \citet{b25} for the degradation of the seeing
measurement, our estimate of the correction may be inaccurate due to
difference in the conditions when the original observations were made.
Regular, long-term monitoring with the same instrument is required to
determine how the site might be changing and whether those changes
could be related to climatic change in Sutherland as has been seen at
other sites \citep{b36}.

\subsection{Site Comparison}
\label{sec:compare}
\begin{table*}
 \centering
 \caption{Comparison of Sutherland with other sites.}
 \label{othersites}
\begin{tabular}{@{}l||c||c|c|c||c|c|l@{}}
  \hline
      & altitude [m] & \multicolumn{3}{c}{median seeing} & $\theta _0$ & $\tau _0$ & \# nights / \# data \\
 Site & & overall & GL & FA & & (MASS$_{corr}$+GL) \\
  \hline
 \bf{Sutherland} & \bf{1768} & \bf{1.4"} & \bf{1.28"} & \bf{0.41"} & \bf{1.92"} & \bf{2.9 ms} & 91 / 29,182  \\
  \textit{March 2010 - March 2011} \\
 Cerro Tololo \citep{b4} & 2207 & 0.88" & 0.44" & 0.50" & 1.56" & 2.9 ms & NA / 433,162 \\
  \textit{2004-2008} \\
 Mauna Kea \citep{b5} & 4050 & 0.71" & 0.51" & 0.42" & 2.69" & 5.1 ms & 124 / NA \\
  \textit{2006-2007} \\
 Paranal \citep{b31} & 2635 & 1.1" & 0.86" & 0.56" & 2.6" & 3.6 ms & NA / NA \\
  \textit{2007} \\
 Mount Graham \citep{b48} & 3221 & 0.95" & 0.81" & 0.39" & 2.5" & 4.8 ms & 43 / 16,659 \\
  \textit{2007}\\
 Cerro Las Campanas \citep{b52} & 2551 & 0.8" & 0.58"$^{\dagger}$ & 0.47" & 1.84" & 2.46 ms$^{\ddagger}$ & 46 / 3,412 \\
  \textit{2010} 
 \\
 \\
 E-ELT \citep{b47} \\
  \textit{2008-2009 ($\sim 1$ year per site)} \\
  Aklim (Morocco) & 2350 & 1.0" & 0.77" & 0.52" & 1.29" & 3.5 ms & NA / 10,992 \\
  Cerro Ma\'{o}n (Argentina) & 4653 & 0.87" & 0.51" & 0.66" & 1.37" & 3.4 ms & NA / 29,723 \\
  Roque de los Muchachos (Canary Island) & 2346 & 0.8" & 0.65" & 0.32" & 1.93" & 5.6 ms & NA / 47,328 \\
  Cerro Ventarrones (Chile) & 2837 & 0.91" & 0.6" & 0.55" & 1.96" & 4.9 ms & NA / 56,547
 \\
 \\
 TMT \citep{b32} \\
  \textit{2003-2008 ($>2.5$ years per site)} \\
  Cerro Tolar (Argentina) & 2290 & 0,63" & 0.34" & 0.44" & 1.93" & 5.2 ms & NA / 196,812 \\
  Cerro Armazones (Chile) & 3064 & 0.64" & 0.35" & 0.43" & 2.04" & 4.6 ms & NA / 212,367 \\
  Cerro Tolonchar (Chile) & 4480 & 0.64" & 0.32" & 0.48" & 1.83" & 5.6 ms & NA / 89,958 \\
  San Pedro M\'{a}rtir (Mexico) & 2830 & 0.79" & 0.58" & 0.37" & 2.03" & 4.2 ms & NA / 139,359 \\
  \hline
  \multicolumn{8}{l}{$\dagger$\textit{GL seeing is not given in arcseconds in the reference paper, hence we calculated it from the MASS and DMM seeing values using the}}\\
  \multicolumn{8}{l}{\textit{following formulae: $see_{GL} = (see_{DIMM}^{5/3} - see_{MASS}^{5/3})^{3/5}$}}\\
  \multicolumn{8}{l}{$\ddagger$\textit{MASS time constant - not corrected}}
 \end{tabular}
\end{table*}
As reported in \S \ref{sec:seeing}, the median integrated seeing
conditions as measured for when both MASS and DIMM were operational
and after a 10 minute binning was 1.4". We use this value to provide
a uniform comparison to other sites, in terms of overall, FA and GL seeing. 
This median seeing is worse than that reported at other major astronomical 
observatories (Table~\ref{othersites}), including Paranal \citep{b31}, Cerro
Tololo \citep{b4}, Mauna Kea \citep{b5}, Mount Graham \citep{b48}, Las Campanas \citep{b52} and both the sites tested for the E-ELT \citep{b47} and TMT \citep{b32}. However, the FA seeing is comparable 
to the best astronomical sites (Table~\ref{othersites}), which confirms 
the fact that turbulence at Sutherland is dominated by the ground layer.
In terms of the other atmospheric parameters, the median isoplanatic 
angle at Sutherland, 1.92", is very similar to most sites. Five of the sites have 
a smaller isoplanatic angle, between 1.29" and 1.84", 3 have comparable values 
between 1.93" and 1.96" and 5 have larger values between 2.03" and 2.69" 
(see \S \ref{sec:iso_time} for details). Despite long coherence time under good 
seeing conditions, 13.7 ms for the best 5\% and 4.9 ms for the best 10\%, 
the median coherence time of 2.7 ms for Sutherland is significantly 
shorter than at other sites that have values between 3.4 ms and 5.6 ms. As seen in equation (4) the GL component of the coherence time is larger 
under weak turbulence (represented by $J_{GL}$). The overall coherence time combines 
both the FA and GL coherence time, but we have seen that the FA turbulence component 
is very similar between sites, hence the coherence time is highly influenced by 
the strong turbulence of the ground layer at Sutherland compare to other sites. However this dependence is also due the method used to derive the 
coherence time from MASS-DIMM data. The accuracy of the coefficient $C$ in equation (4) is questionable. Comparing $\tau _0$ results from MASS and DIMM measurements also shows discrepancies. Previous results from M. Sarazin at Paranal \footnote{http://www.eso.org/gen-fac/pubs/astclim/paranal/asm/mass\\/MASS-Paranal-2003/}, reported in \citet{b49}, show that $\tau _0 ^{DIMM}$, calculated from equation (3) in \citet{b31}, gives values 2.5 larger than $\tau _0 ^{MASS}$. Here the correction we applied gives 
$\tau _0 \sim 1.31 \tau _0 ^{MASS}$, which might still be underestimating the actual 
value. Considering the discrepancy between measurements from MASS and other instruments, as well as the low reliability of the correction applied due to the wide range of value for the coefficient $C$ found in the literature, more investigation on the $\tau _0$ value is needed.

The observed overall worse seeing conditions at Sutherland compared to
other sites can be partly explained due to several discrepancies
between our data set and those from other sites results.  First, many
of these sites have only fully published results from earlier
periods and we are not comparing results over the same time period.
Recent results from Paranal indicated the seeing conditions have
degraded over the last decade and that this may be due to longer term
climatic changes \citep{b36}.

Another important caveat is that the DIMMs at the comparison sites are
typically located on 5 m high towers, while the Sutherland DIMM is
located at ground level. Hence, our results may be strongly affected
by convective turbulence from the ground. Using the results from
MASS-DIMM, SLODAR profiles, and previous studies, we can estimate the
effect due to the lower 5 m surface layer. From MASS-DIMM, we found
that 84\% of the turbulence is located in the GL, and the SLODAR results
tell us that the lowest 30 m contributed 25\% of the GL. In \citet{b12}, 
it is reported that the first 5 m have a seeing of 0.15", which correspond to 66.3\% of the turbulence associated to the 0.19" seeing of the first 30 m for
that period. Using the same proportion we can reasonably consider that
the first 5 m contributes roughly 14\% of the overall turbulence. From
this, we would expect that the integrated median seeing measured from a 5 m
platform would be 1.28". Following the same procedure, the expected
seeing at the dome entrance for SALT, which is located at approximately
30 m, would be 1.22". In addition, when fitting the dome/tube seeing for profile 
reconstruction in SLODAR, part of the turbulence outside the dome belonging to the 
ground layer might be subtracted resulting in an underestimation of both the 
ground layer and the overall turbulence. As a result the first 30 m may be contributing more than 25\% of the turbulence.
\\
However, despite the facts considered above, the altitude of the Sutherland site (1768 m) 
is much lower than all the other sites whith altitudes ranging from 2290 m to 4653 m, 
so we do not expect the seeing at Sutherland to be as good as these other sites.

In terms of GL, both \citet{b5} and \citet{b48}, respectively for Mauna Kea and Mount Graham, presented high vertical resolution turbulence profiles. The Sutherland site differs signifcantly from those two sites due to its strong upper ground layer located at 300 to 500 m. The Sutherland upper GL contributes nearly 50\% of the GL turbulence while it is only 2\% and 12\% for Mauna Kea and Mount Graham, respectively. Most of the turbulence at those sites is located in the lower layers, 90\% in the first 40 m for Mauna Kea and 70\% in the first 100 m for Mount Graham, compare to 30\% in the first 100 m at Sutherland.
\subsection{Weather conditions bias}
\begin{figure*}
\includegraphics[scale=0.65]{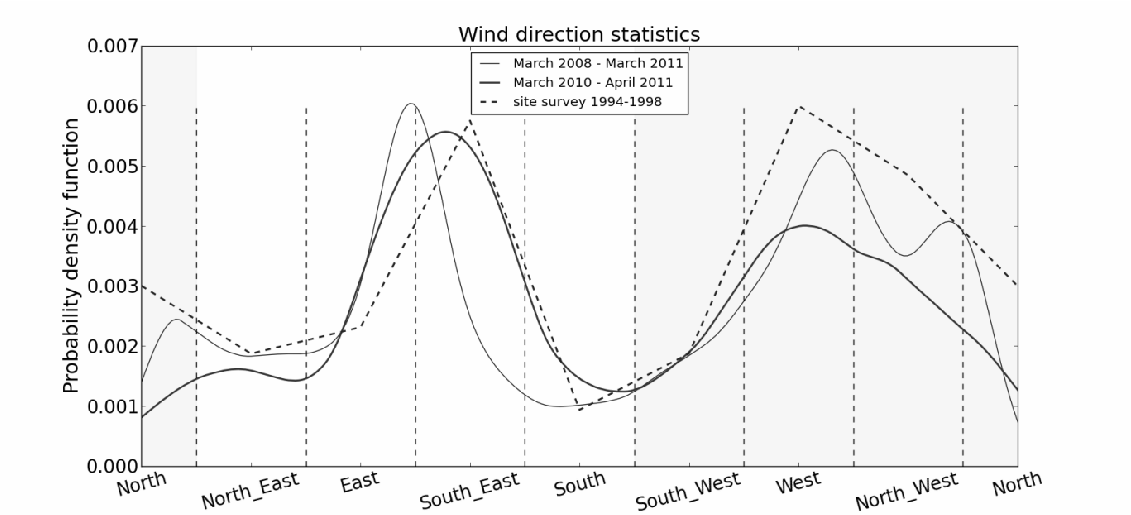}
\caption{Wind direction and seeing.  Wind direction frequencies during
  our seeing measurements (thick line) compare to those of the
  past 3 years (thin line) and those of the site testing
  campaign \citep{b12} (dashed line). The x axis indicates the
  wind directions. The shaded area correspond to wind directions
  associated with good seeing.}
 \label{wind3yr}
\end{figure*}

In order to comment on potential bias of our data set due to weather
conditions during our observing period, we used weather data from the
SALT weather station taken between April 2008 and April 2011 and
consisting of 966 nights and 729663 data points. A comparison between
the wind directions during this period and our site monitoring
campaign indicates that the typical seeing at Sutherland is expected
to be lower than the results presented here. For the following discussion
we will use all DIMM measurements without any averaging.

Comparing the occurrence of the wind directions during our site survey
campaign with three years of weather data (presented in Fig.
~\ref{wind3yr}), one would expect the seeing to be slightly better
over this three year period as conditions associated with good seeing
(winds from the westerly direction) are more prevalent than during
our site monitoring campaign. Indeed, during our campaign, winds
associated with bad seeing correspond to 39.4\% of our data while those
associated with good seeing correspond to 49.2\%. In comparison, the long term weather data indicate winds
associated to bad seeing occur only 28.3\% of the time, while those associated to good seeing represent 58.4\% of the
data.  Moreover, the wind pattern for the past three years is in very
good agreement with that reported in \citet{b12} (dashed line in Fig.
\ref{wind3yr}), which showed good seeing conditions 57\% of the time
and bad seeing conditions 29.4\% of the time.

If we use the past three years as representing the typical wind
pattern, we can estimate the median seeing conditions for Sutherland
based on the distribution of wind direction and the correlation
between the seeing and wind direction presented in \S \ref{res:wind}.
Using these values and the three year wind direction data, we expect
the median seeing to be around 1.2" as compared to that measured for
the year 2010-2011, including all DIMM data without any averaging, of 1.32".

\section{Conclusions}
While previous studies of the Sutherland site provided a good general
overview of the observing conditions, they were lacking the turbulence profile
and did not reflect the current conditions.  Hence the present
study provides an up-to-date view of the Sutherland site seeing
conditions.  In addition, it also provides a detailed monitoring of
the atmospheric turbulent profile, albeit only over one year.

This first year of seeing monitoring resulted in the successful
installation of a fully automated MASS-DIMM providing continuous
seeing monitoring for all observers on the plateau at Sutherland. With
a year of seeing measurements consisting of 27 days of SLODAR
operations, 91 days of MASS-DIMM data, and 55 additional days of TimDIMM
measurements, we obtained the first atmospheric turbulence measurements
at the Sutherland site.

The median seeing for the Sutherland site from these data is 1.4".
Like most astronomical sites, Sutherland has 3 main layers
contributing to the turbulence: the GL below 1 km, a layer in which
turbulence is driven by the wind shear whose altitude varies between 2
and 5 km, and finally the upper layers above 12 km where turbulence
are associated to the jet-stream. Under median conditions, the GL
contributes 84\% of the turbulence of the integrated seeing. Moreover,
we know from SLODAR profiles that the GL is dominated by the first 30 m
and the higher 300 to 500 m.

With this instrumentation in place and a future upgrade with a 5 m
tower, long-term monitoring of the site will allow further
investigation of any seasonal bias to atmospheric conditions, as well
as longer timescale climatic effect, e.g. associated with El
Ni\~{n}o/La Ni\~{n}a. The availability of real-time seeing data, along
with the behaviour of the atmospheric conditions, will allow for
improved observing planning efficiencies, especially at SALT, which is
a queue scheduled telescope. The preponderance of the GL turbulence is 
promising for the development of a ground layer adaptive optics systems 
for the Sutherland telescopes, and future campaigns with additional 
site testing equipment will refine the turbulence profile measured here, 
as well as the coherence time value.

\section{Acknowledgements}
This work is based upon research supported by the National Research Foundation (NRF) and the UK Science \& Technology Facilities Council.\\
The authors are thankful to the referee for the valuable and constructive comments and suggestions given. We also wish to thank the SALT Foundation, who funded and own the seeing instruments, as well as the SAAO and SALT operations staff for the logistical support in this study.
\bsp
\label{lastpage}
\end{document}